\documentclass[journal=jacsat,manuscript=article]{achemso}

\usepackage[utf8]{inputenc}
\usepackage[T1]{fontenc}

\usepackage{graphicx}
\usepackage{color}

\usepackage{graphicx}
\usepackage[percent]{overpic}

\usepackage[version=3]{mhchem} 

\author{Sofia Zinzani}
\affiliation{Università degli Studi di Milano, Physics Department, Milano, IT}

\author{Francesca Baletto}
\affiliation{Università degli Studi di Milano, Physics Department, Milano, IT}

\author{Kevin Rossi}
\affiliation{TU Delft, Materials Science and Engineering Department, Delft, NL}
\alsoaffiliation{TU Delft | The Hague, Climate Safety and Security Centre, The Hague, NL}
\email{k.r.rossi@tudelft.nl}

\title[An \textsf{achemso} demo]
{Bridging Structure and Activity in Nanocatalysts via Machine Learning and Global Structure Representations}


\begin{document}

\begin{tocentry}

\centering
\includegraphics[width=1\linewidth]{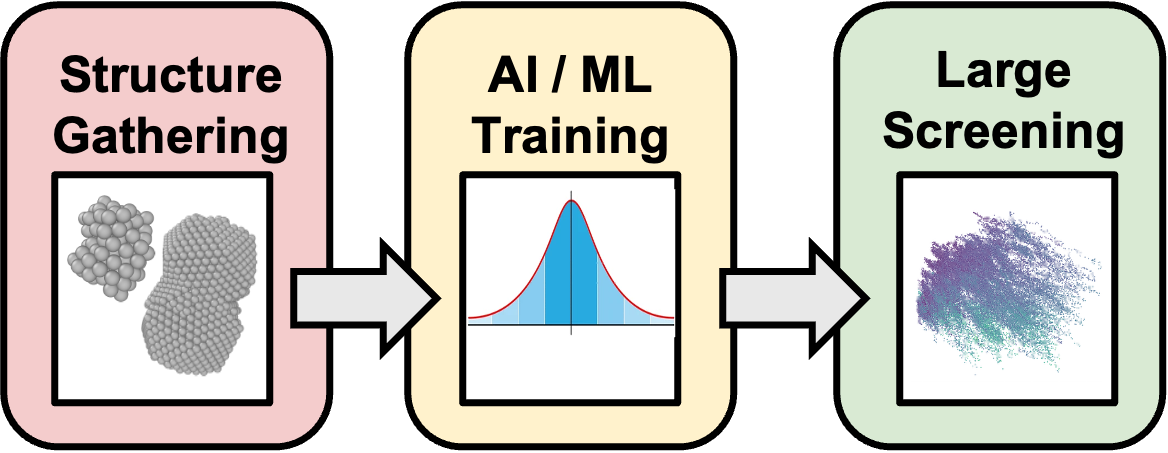}





\end{tocentry}

\begin{abstract}
Establishing a mapping between nanocatalysts structure and their catalytic properties is essential for efficient design.
To this end, we demonstrate the accuracy of a general machine learning framework on a representative and challenging application: predicting the mass activity of Pt nanoparticles for the electrochemical oxygen reduction reaction, estimated via a microkinetic model.
Accurate models are obtained when leveraging either a nanocatalyst's structure representation accessible at the computational level, namely the surface site generalized coordination number distributions, or one accessible experimentally, namely the nanoparticle's pair distance distribution function.
Building on this result, we demonstrate that our machine learning model, in tandem with Bayesian optimization, efficiently identifies the Top-10 and Top-100 most active structures out of a large pool of candidates comprising more than 50000 different structures, after probing the activity only of a few thousand structures.
These findings provide a robust blueprint for accelerated theoretical and experimental identification of active nanocatalysts.
\end{abstract}


The development of next-generation heterogeneous catalysts with enhanced activity, selectivity, and stability directly underpins the large-scale deployment of clean energy solutions, the production of value-added chemicals, and the mitigation of carbon emissions.\cite{Friend2017}
Indeed global sustainability and strategic autonomy require transformative advances in the design of heterogeneous catalysts, which currently appear in at least one step of 80\% industrial processes.\cite{Rothenberg2008} 
In this context, nanocatalysts provide a versatile platform for tuning catalytic performance through controlled manipulation of their size, shape, chemical composition and ordering.\cite{Mitchell2021, Mistry2016}
Such flexibility allows researchers to optimize  surface electronic structures, surface sites adsorption properties, and active site density, to boost reactivity and selectivity in a wide range of applications.

Machine Learning (ML) models can learn intricate structure–reactivity correlations from simulation-generated or experiments and provide rapid performance predictions for new structural or compositional inputs.\cite{Esterhuizen2022, Foppa2021}
ML applications have obtained remarkable successes in predicting intermediate-scale properties - such as adsorption energies or reaction barriers, which are crucial descriptors in catalytic processes.\cite{PabloGarcia2023FastAdsorptionGNN, Chen2025MultimodalTransformer, Lan2023AdsorbML} 
Establishing a reliable and direct map between morphological features of catalysts and macroscopic performance metrics - such as the overall catalytic activity - has remained elusive, in so far.\cite{Mauss2025}

In this work, we demonstrate a general ML workflow capable to address this gap.
As an illustrative case study, we apply it to learn and evaluate the relationship between the structure of platinum (Pt) nanoparticles (NPs) and their catalytic activity for the electrochemical oxygen reduction reaction (eORR)—a key process in green energy technologies such as fuel cells and electrolyzers.\cite{Braunwarth2020}
Given the high cost and limited availability of Pt, it is indeed desirable to identify nanocatalyst geometries that maximize catalytic performance per unit mass.

\begin{figure}[h]
    \centering
    \includegraphics[width=0.95\textwidth]{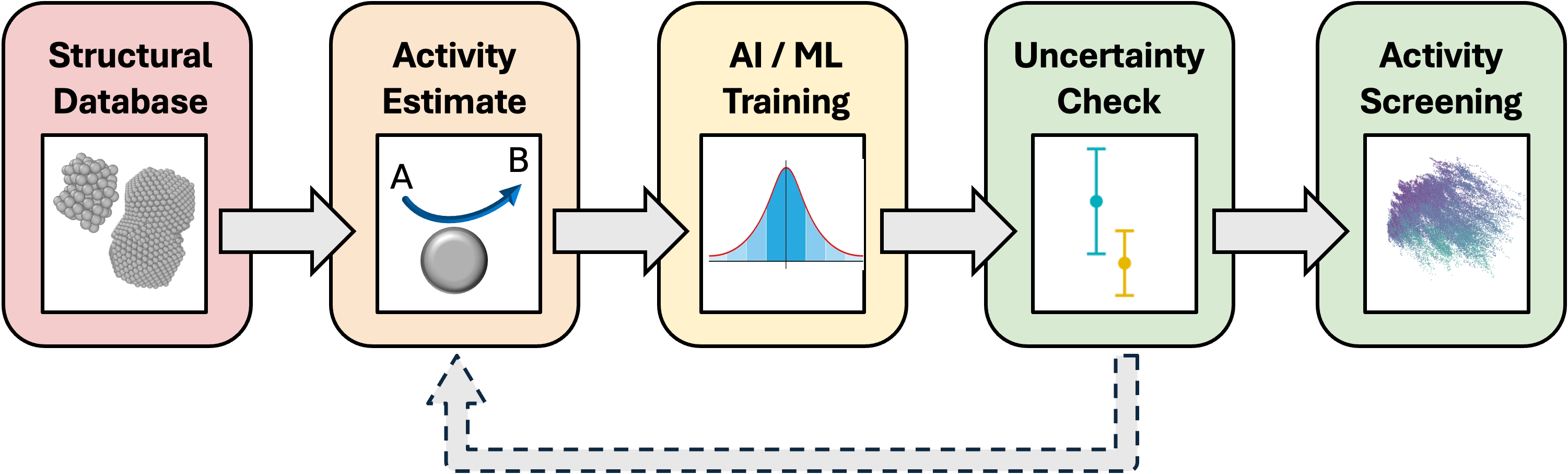}
    \caption{Graphical illustration of our workflow to bridge the modeling gap in heterogeneous catalysis. First we generate an extensive set of structures containing a diversity of non equivalent active sites distributions. We evaluate their activity for a specific reaction and product. We train a machine learning algorithm to chart structure-activity relationship and rapidly (fraction of a second) offer prediction on the catalytic properties of a large number of other structures.}
    \label{fig:fig1}
\end{figure}

The workflow, illustrated in Figure \ref{fig:fig1}, consists of five stages.
In first instance, we generate a large database of 52318 structures sampled during previous numerical investigations (See SI - Dataset Generation for further detail).
This dataset comprises unique nanoparticle structures with sizes ranging from 264 to 2,830 atoms, corresponding to approximately 2 to 6 nm.
Next, we evaluate the nanoparticles mass activity for eORR at an applied voltage of 0.9V (MA$_{@0.9V}$ [A/mg]).
To estimate this quantity, we adopt a previously discussed and validated microkinetic model \cite{Rossi2020} where surface site contributions to the overall nanoparticle activity for eORR are estimated as a function of their atop generalized coordination number (GCN) \cite{CalleVallejo2015_NatChem, CalleVallejo2015_Science} (see SI - Structure-Activity Theory and Modeling for further detail).

Since we expect the workflow to be applied in a low-data regime, we adopt a Gaussian Process Regression (GPR)\cite{rasmussen2006gaussian} to learn the mapping between geometrical features of the nanoparticle structure and its activity.
To test the ML model, we consider an ideal representation of the NP surface, accessible at the computational level, that is the GCN distribution of atop sites.
We then demonstrate the workflow generality and effectiveness considering an experimentally accessible representation of the nanosystem, namely its pair distance distribution function (see SI - Gaussian Process regression for further detail)

The uncertainty in the GPR model prediction is used either in an active learning or in a Bayesian optimization \cite{frazier2018tutorial} loop, to intelligently augment the training datapoints and improve the model accuracy in regions of interest, with the end goal of screening a large number of geometries, and identifying the most active ones (see SI - Data-efficient Learning section for further methodological details).

\begin{figure}
    \centering
    \includegraphics[width=0.45\linewidth]{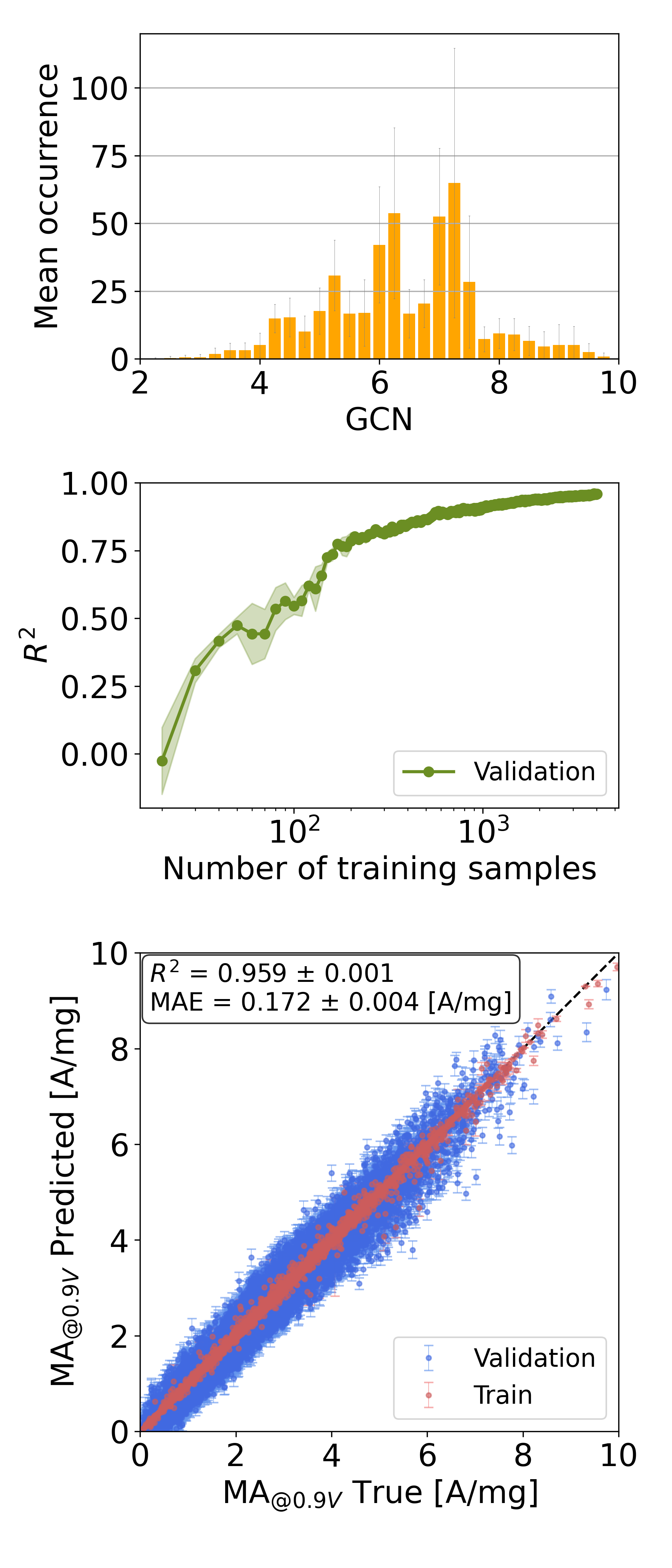}
    \caption{
        Top) Mean GCN distribution. To estimate this quantity we average the GCN distribution for each isomer in the dataset. The GCN distribution standard deviation, when considering each isomer in the dataset, is represented by the errorbars. 
        GCN distribution bins have a resolution of 0.25.
        Mid) GCN  Learning curve observed when training a machine learning model to predict Pt nanoparticle activity for electrochemical oxygen reduction reaction from nanoparticle generalized coordination number distributions. Bottom) Corresponding parity plot between predicted and true activities for machine learning model trained with 4020 datapoints selected via an active learning protocol. 
    }
    \label{fig:fig2}
\end{figure}

As an initial proof of concept, we discuss the application of this workflow when adopting the GCN as the nanoparticle structure descriptor. 
The initial database of 52318 structure results in diverse GCN distributions (binned with a resolution 0.25) as illustrated by the histograms in the top panel of Figure \ref{fig:fig2}. 
Because of our binning resolution, GCN signature at at 6.5 < GCN $\leq$ 6.75 and
7.25 < GCN $\leq$ 7.5 
correspond to the presence of (100) and (111) facets, respectively. 
Atoms with GCN > 8 instead correspond to convex-sites, where a GCN ~8.33 was demonstrated to correspond to the most active Pt sites for eORR.\cite{calle-vallejo2017why}

To analyse the relationship between model accuracy and training points samples, we evaluate the GPR model learning curve, illustrated in Figure \ref{fig:fig2}, mid panel.
The learning curve is evaluated by iteratively adding training samples to an initial randomly selected set of 10 data points, with each new sample chosen according to an active learning approach.
A significant improvement in the model accuracy is witnessed when performing active learning until 500 samples are present in the training set ($R^2=0.865\pm0.003$).
%
%
The model accuracy further increases, albeit with a slower pace, also with additional training samples, almost reaching saturation.

After 400 active learning iterations, corresponding to 4020 training points (last point in Figure \ref{fig:fig2} central panel)
the model achieves near-perfect accuracy, with a correlation coefficient of $R^2=0.959\pm0.001$ and a Mean Absolute Error (MAE)$=0.172\pm0.003~ A/mg$ between predicted and reference validation mass activity values (Figure \ref{fig:fig2} bottom panel, where training points are reported in red and validation point in blue).
Such an high accuracy is achievable since the ground truth data are obtained via a microkinetic model that utilizes the catalyst surface sites GCN to estimate MA$_{@0.9V}$ [A/mg]. 
Also, according to intuition, the accuracy of the model prediction is found to increase or decrease depending on whether the GCN distribution is described with a finer or more coarse binning
(SI - Figure S5).

Building upon the promising results achieved using the GCN distribution as a global structural representation, we investigate an alternative one: the nanoparticle pair distance distribution function (PDDF).
The motivation for considering this representation lies in its proven connection to experimental observables. PDDFs can be obtained from total scattering measurements using X-rays, neutrons, or electrons.\cite{terban2022structural}
Further in a previous work, we demonstrated how PDDFs track melting signatures and surface-bulk ordering in metallic nanoparticles.\cite{delgado2021universal}
More recently, the use of nanoparticle PDDFs to map structural heterogeneity in solid nanoparticles has been demonstrated too.\cite{telari2023charting} 
Furthermore, experimentally, PDDF analyses of Pt-based nanostructures revealed that microstrain correlates with enhanced eORR kinetics.\cite{chattot2017beyond,Chattot2020} 
Last but not least, approaches to reconstruct coordination number distributions from XAS spectra have been suggested.\cite{timoshenko2017supervised}

\begin{figure}
    \centering
    \includegraphics[width=0.45\linewidth]{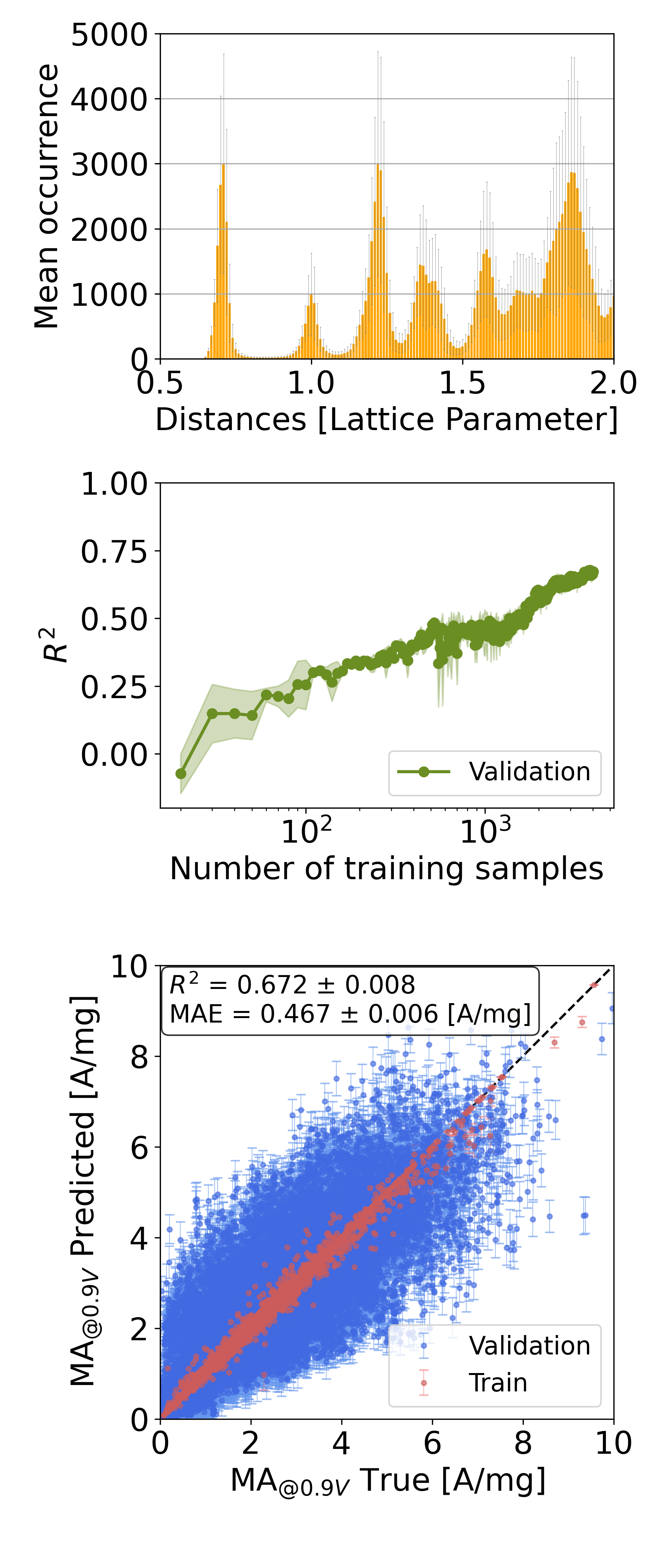}
    \caption{
        Top) Mean PDDF values distribution over the full dataset of structures with its uncertainty represented by the errorbars. PDDF bins have a resolution of 0.01 lattice parameter. 
        Mid) PDDF  Learning curve observed when training a machine learning model to predict Pt nanoparticles activity for eORR.
        Bottom) Corresponding parity plot between predicted and true activities for machine learning model trained with 4020 datapoints selected via an active learning protocol. 
    }
    \label{fig:fig3}
\end{figure}

Our findings, summarized in Figure \ref{fig:fig3}, provide fair evidence supporting the hypothesis that machine learning models can effectively utilize the PDDF as a representation to predict nanocatalysts' activity.
As illustrated in Figure \ref{fig:fig3} top panel, 
we focus on interatomic pair distances up to a cut-off distance of twice the bulk lattice parameter.
In a crystalline FCC bulk, first and second nearest-neighbors, at $\sqrt{2}/2$ and 1 lattice distances, respectively, encode information about atoms first and second coordination shells.
Deviation from ideal bulk lattice value in these distances, and beyond, encode bond distortions due to specific geometrical feature of the NP e.g., grain boundaries, strain, and the presence of low coordination atoms or adatoms islands.

The learning curve depicted in Figure \ref{fig:fig3} mid panel demonstrates that a sizable but not intractable training set, comprising a few thousand samples selected via active learning, enables an ML model with a fair accuracy. 
A $R^2$ over 0.5 is indeed achieved with more than 2000 training samples.
Notably, we do not observe a saturation in the model accuracy when increasing  the number of training samples, within the training set sizes here considered (up to 4020 training samples).


For the representative case where the model is trained on 4020 samples, it achieves a good predictive performance, with a correlation coefficient of $R^2=0.672\pm0.008$ and a mean absolute error (MAE) of $0.467\pm0.006 ~ A/mg$ between the predicted and reference values for mass activity, as shown in Figure \ref{fig:fig3} bottom panel.
Reported trends hold for different choices of pair distance distribution function binning and maximum distances considered (SI - Figure S6-S11).

While the results show that GPR models trained on the PDDF representation achieve reasonably accurate predictions across a broad range of structures, (arguably) catalyst design is ultimately concerned with prioritizing the discovery of top-performing candidates. 
Thus, a key question remains whether this approach would enable to reliably identify most catalytically active materials in a large pool of candidates.

To address this question, we explore the use of Gaussian Process Regression leveraging the PDDF representation and Bayesian optimization (BO) \cite{frazier2018tutorial}, to efficiently pinpoint the most active Pt nanoparticles for eORR in our dataset.
To guide candidate selection in our BO, we adopt an Upper Confidence Bound (UCB) acquisition function.\cite{srinivas2012information} The latter is defined as
$ \text{UCB}(x) = \mu(x) + \kappa \cdot \sigma(x) $,
where \(\mu(x)\) and \(\sigma(x)\) denote the predictive mean and standard deviation of the surrogate model at point \(x\) and \(\kappa\) is a tunable hyperparameter that controls the exploration–exploitation tradeoff.
In our application, \(\mu(x)\) and \(\sigma(x)\) label the predicted MA$_{@0.9V}$ [A/mg] and the uncertainty on this prediction, for a given nanoparticle structure, respectively.
We report data with $\kappa$ =2, which is empirically observed to strike the best balance between exploration and exploitation. 
Larger $\kappa$ (e.g, $\kappa$ =10) result in less efficient sampling of highly active areas, while lower $\kappa$ (e.g., $\kappa$ =0.2) get stuck in exploring only specific areas (SI - Figure S12).

\begin{figure}
    \centering 
    \includegraphics[width=0.99\linewidth]{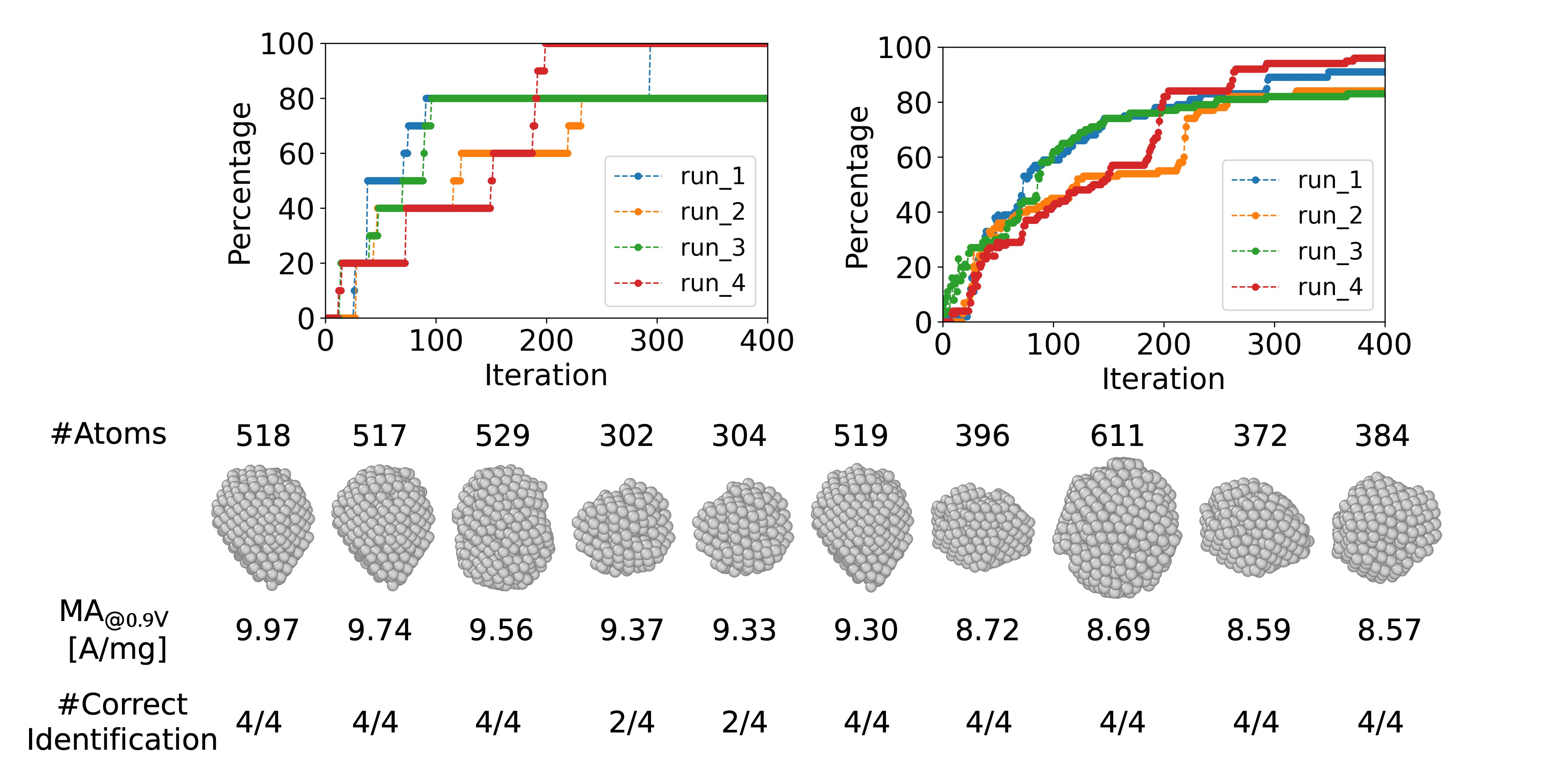}
    \caption{ Top) Percentage of predicted points, per iteration, that match the values of the top 10 (left) and top 100 (right) true maxima of MA$_{@0.9V}$ [A/mg]. Bottom) Visual illustration of the nanoparticles displaying the top-10 largest MA$_{@0.9V}$ [A/mg]. For each we report the system size (\#atoms), the MA$_{@0.9V}$ [A/mg], and the relative amount of times the four Bayesian optimization runs correctly identify them among the top-10 most active structures. }
    \label{fig:fig4}
\end{figure}

As illustrated in Figure \ref{fig:fig4}, starting from a modest initial training set of 10 randomly selected samples, our BO efficiently navigates the candidate pool of 52318 structures to swiftly identify the most active structures, 
notwithstanding, small changes in nanoparticle morphology induce a large change in MA$_{@0.9V}$ [A/mg].
Within just 200 acquisition steps - where 10 structures from the candidate pool are selected at each acquisition step  - the algorithm successfully identifies more than half of the top 100 and top 10 (Figure \ref{fig:fig4} lower panel) most active nanoparticles. 
With 300 acquisition steps, at least 80\% of the top-10 and top-100 structures are correctly identified, where misclassification are observed only for the case of the smallest highly active nanoparticles, with sizes around 300 atoms.
While more systematic assessments would be necessary to rationalize this behavior, we hypothesize that this shortcoming stems from the fact that smaller nanoparticles present unique pair distance distortions due to finite size effects.

In conclusion, we present a machine learning framework capable of quantitatively capturing structure–activity relationships, guiding high-throughput screening by leveraging structural representations accessible at a theoretical level, i.e., generalized coordination number distribution, and experimentally, i.e., pair distance distribution.

Accurate machine learning predictions of structure–activity relationships were expected when representing nanoparticles using the distribution of atop generalized coordination numbers of their surface sites, since these same were used as inputs in the microkinetic model that generated the ground truth data.
For the case of a PDDF representation, the literature, e.g., including but not limited to \cite{chattot2017beyond,Chattot2020}, has shown that this quantity is of aid to qualitatively rationalizing catalytic trends.
Through our machine learning approach, we propose that PDDF measurements can be also used for quantitative prediction and identification of most active catalytic structures among a large pool of candidates.

While we validate the framework computationally -- similar to other efforts connecting experimentally and theoretically derived quantities relevant to catalysis \cite{wang2022quantitatively, lansford2020infrared} -- we emphasize that our approach could be readily integrated into (autonomous) experimental platforms, where high-throughput synthesis and characterization support iterative, on-the-fly, model refinement.\cite{anker2025autonomous} 
Beyond the specific application shown here, our approach is also compatible with other modeling paradigms for estimating catalyst activity and selectivity, such as kinetic Monte Carlo simulations, making it a versatile tool for accelerating catalyst discovery.
Indeed, we expect our approach to be general across diverse reactions and systems. 
Future works can thus also consider more complex catalytic systems, including multi-element materials and supported architectures, and more complex reactions.

\section{Author Information}

Corresponding Author \\
Kevin Rossi - Materials Science and Engineering Department, Delft University of Technology, Delft, 2623CD, Netherlands - ORCID:0000-0001-8428-5127 ; email: k.r.rossi@tudelft.nl \\
Authos\\
Francesca Baletto - Department of Physics, University of Milan, Milan, 20133, Italy - ORCID: 0000-0003-1650-0010 \\
Authors \\
Sofia Zinzani - Department of Physics, University of Milan, Milan, 20133, Italy - ORCID:
0009-0009-6961-7766

\section{Author contributions}
K.R. conceived the project(s).
S.Z. and K.R. conducted the experiment(s). 
All authors analysed the results and reviewed the manuscript.

\section{Conflicts of interest} 
There are no conflicts to declare.

\section*{Data availability}
.xyz coordinates of the Pt nanoparticles considered in the work, together with their GCN and PDDF are available upon acceptance at: *zenodo link*.
Python codes for GPR structure-activity prediction, active learning, and Bayesian Optimization are available at: \url{https://github.com/sofiazinzani/ML_GCN_PDDF_ORR}.
The Git repository also contains all the outputs of the codes when executed, including scripts to reproduce the graphs reported in this manuscript.
Python codes for GCN and PDDF calculations are available at : \url{https://github.com/nanoMLMS/pySNOW}.

\section{Acknowledgements}

This article is based upon work from COST Action CA22154 - Data-driven Applications towards the Engineering of functional Materials: an Open Network (DAEMON) supported by COST (European Cooperation in Science and Technology).
SZ acknowledges the University of Milan and ISC SRL for financially supporting her PhD studentship (D.M. 118/2023 PNRR).  
\begin{suppinfo}

The Supporting Information is available free of charge at XXX.

\end{suppinfo}

\bibliography{acs-bibliography}

\end{document}